# Frequency domain TRINICON-based blind source separation method with multi-source activity detection for sparsely mixed signals

Zelin Wang
*Key Laboratory of Modern Acoustics and Institute of Acoustics*
*Nanjing University*
Nanjing, People's Republic of China
zelinwang@smail.nju.edu.cn

Jing Lu
*Key Laboratory of Modern Acoustics and Institute of Acoustics*
*Nanjing University*
Nanjing, People's Republic of China
lujing@nju.edu.cn

Kai Chen
*Key Laboratory of Modern Acoustics and Institute of Acoustics*
*Nanjing University*
Nanjing, People's Republic of China
chenkai@nju.edu.cn

*Abstract*—The TRINICON ('Triple-N ICA for convolutive mixtures') framework is an effective blind signal separation (BSS) method for separating sound sources from convolutive mixtures. It makes full use of the non-whiteness, non-stationarity and non-Gaussianity properties of the source signals and can be implemented either in time domain or in frequency domain, avoiding the notorious internal permutation problem. It usually has best performance when the sources are continuously mixed. In this paper, the offline dual-channel frequency domain TRINICON implementation for sparsely mixed signals is investigated, and a multi-source activity detection is proposed to locate the active period of each source, based on which the filter updating strategy is regularized to improve the separation performance. The objective metric provided by the BSSEVAL toolkit is utilized to evaluate the performance of the proposed scheme.

## I. INTRODUCTION

The problem of separating convolutive mixtures of unknown time series arises in many fields. A prominent example is the so-called cocktail party problem [1], where individual speech signals are expected to be extracted from mixtures of multiple speakers in a reverberant acoustic environment. Blind source separation (BSS) aims to solve this problem usually with the assumption that signals of different sound sources are statistically independent from each other [1].

The well-known independent component analysis (ICA) [2] is an efficient BSS approach, which is often realized in frequency domain to deal with convolutive mixtures. The internal permutation ambiguity of the frequency domain ICA requires a proper supplementary repair mechanism based on the information of inter-frequency correlations or direction of arrivals (DOA) [3]. The internal permutation can also be resolved by extending the ICA to multivariate case, yielding independent vector analysis (IVA) [4], or the TRINICON ('Triple-N ICA for convolutive mixtures') method [5-6]. The TRINICON method can be efficiently implemented in frequency domain using the second-order statistics (SOS) [7-8], and is of particular interest in this paper.

The offline TRINICON method usually has the best performance when the sources are continuously mixed. However, in practical applications, speech signals are sometimes sparsely mixed, leading to possible performance deterioration because all the signal segments are equally evaluated in the algorithm regardless of the source activity.

In this paper, the implementation of the frequency domain offline TRINICON algorithm is investigated and a straightforward but effective multi-source activity detection is proposed to approximately identify the activity of each source. The filter updating process is regularized according to the source activity detection results, and the improved separation performance for sparsely mixed signals are demonstrated through simulations. Exemplary audio samples are available online at https://github.com/zelinwangnju/FB_TRINICON_withAD.

## II. PROBLEM FORMULATION

Fig. 1 shows the signal model of the dual-channel BSS system, where $s_1$ and $s_2$ in Fig. 1 represent two sound sources, $x_1$ and $x_2$ are the signals received by the microphones, $y_1$ and $y_2$ are the output signals, $h_{up}$ is the room transfer function from the $u$th source to the $p$th microphone, and $w_{pq}$ denotes the demixing filter from the $p$th input signal to the $q$th output signal. In Fig. 1(a), two sources are located in two half planes separated by the midline of the array, where causal filters are sufficient to achieve interference cancellation. However, if the sources are located in the same half plane as shown in Fig. 1(b), one noncausal filters $w_{12}$ or $w_{21}$ are required.

As introduced by [5] and [9], the cost function of TRINICON is defined based on the generalization of Shannon's mutual information as:

$$\mathcal{J}(m) = \sum_{i=0}^{\infty} \beta(i,m) \frac{1}{N} \sum_{j=0}^{N-1} \left\{ \log\left( \hat{p}_{QD}\left( \mathbf{y}_1(i,j),...,\mathbf{y}_Q(i,j) \right) \right) \right.$$
$$\left. - \log\left( \hat{p}_{1,D}\left( \mathbf{y}_1(i,j) \right) \cdot ... \cdot \hat{p}_{Q,D}\left( \mathbf{y}_Q(i,j) \right) \right) \right\}, \quad (1)$$

where $\hat{p}_{q,D}\left( \mathbf{y}_q(i,j) \right)$ is the estimated multivariate probability density function (PDF) for channel $q$ of dimension $D$,

This work was supported by the National Natural Science Foundation of China with Grant No. 11374156.

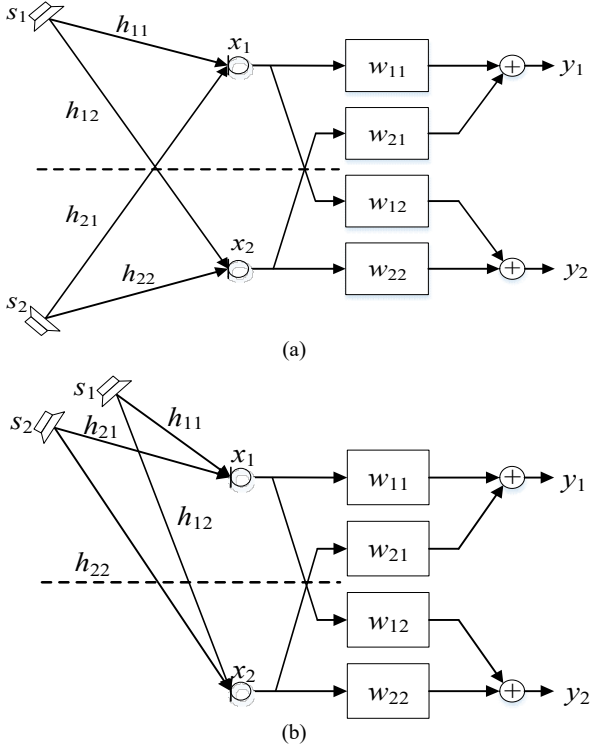

Fig. 1. The signal model of the BSS system with (a) only causal delays and (b) causal and noncausal delays for the demixing system

$\hat{p}_{QD}(\mathbf{y}_1(i,j),...,\mathbf{y}_Q(i,j))$ is the estimate of the joint PDF of dimension $QD$ over all $Q$ (here, $Q = 2$) output channels, $m$ is the block index, $j = 0, …, N − 1$ is the time-shift index within a block of length $N$, and $\beta(i,m)$ is a weighting function allowing offline and online implementations of the algorithm, normalized according to $\sum_{i=0}^{\infty}\beta(i,m) = 1$.

Exploiting the second order statistics (SOS) [7][10], the cost function is expressed by

$$\mathcal{J}(m) = -\sum_{i=0}^{\infty}\beta(i,m)\log\left(\frac{\det \mathrm{bdiag}\left[\mathbf{Y}^{\mathrm{H}}(i)\mathbf{Y}(i)\right]}{\det \mathbf{Y}^{\mathrm{H}}(i)\mathbf{Y}(i)}\right), \quad (2)$$

where the bdiag operation on a partitioned block matrix sets all submatrices on the off-diagonals to zero

The filter matrix $\mathbf{W}$ is updated by the natural gradient descent [11] method as

$$(\mathbf{W}(m))^{(j_{iter})} = (\mathbf{W}(m))^{(j_{iter}-1)} - \mu \nabla_{\mathbf{W}(m)}^{NG}\mathcal{J}(m), \quad (3)$$

where

$$\nabla_{\mathbf{W}}^{NG}\mathcal{J}(m) = 2\sum_{i=0}^{\infty}\beta(i,m)\mathbf{W}\{\mathbf{Y}^{\mathrm{H}}(i)\mathbf{Y}(i) \\ - \mathrm{bdiag}[\mathbf{Y}^{\mathrm{H}}(i)\mathbf{Y}(i)]\}\mathrm{bdiag}^{-1}[\mathbf{Y}^{\mathrm{H}}(i)\mathbf{Y}(i)], \quad (4)$$

with $j_{iter}$ the iteration number and $\mu$ the step size.

For the frequency domain implementation [7], the filter matrix and the signals are expressed as (the block index $m$ is omitted for brevity and the underbar denotes the parameter in frequency domain)

$$\underline{\mathbf{W}}_{pq} = \mathrm{diag}\left\{\mathbf{F}_{4L\times 4L}\left[w_{qu,0},...,w_{qu,L-1},0,...,0\right]^{\mathrm{T}}\right\} \\ \underline{\mathbf{W}} = \begin{bmatrix} \underline{\mathbf{W}}_{11} & \underline{\mathbf{W}}_{12} \\ \underline{\mathbf{W}}_{21} & \underline{\mathbf{W}}_{22} \end{bmatrix} \quad (5)$$

and

$$\underline{\mathbf{X}}_p = \mathrm{diag}\left\{\mathbf{F}_{4L\times 4L}\left[x_p(mL-3L),...,x_p(mL+L-1)\right]^{\mathrm{T}}\right\} \\ \underline{\mathbf{X}} = [\underline{\mathbf{X}}_1, \underline{\mathbf{X}}_2] \quad (6)$$

respectively, where $L$ is the length of the filter and $\mathbf{F}_{4L\times 4L}$ is the Fourier transform matrix.

The power spectral density (PSD) matrix $\mathbf{\Phi}_{yy}$ and $\mathbf{\Phi}_{xx}$ are expressed as

$$\mathbf{\Phi}_{yy} = \underline{\mathbf{W}}^{\mathrm{H}}\mathbf{G}_{4LP\times 4LP}^{1_{2L}0}\mathbf{\Phi}_{xx}\mathbf{G}_{4LP\times 4LP}^{1_{2L}0}\underline{\mathbf{W}} \quad (7)$$

and

$$\mathbf{\Phi}_{xx} = \underline{\mathbf{X}}^{\mathrm{H}}\mathbf{G}_{4L\times 4L}^{01_L}\underline{\mathbf{X}}, \quad (8)$$

respectively. Matrix $\mathbf{G}$ with different dimensions are mainly responsible for preventing decoupling of the individual frequency components and thus avoiding the internal permutation and circular convolution effects [5]. The gradient in frequency domain can be expressed as [12-13]:

$$\nabla_{\underline{\mathbf{w}}^{(k)}}^{NG}\mathcal{J}^{(k)} = 2\sum_{i=0}^{\infty}\beta(i,m)\underline{\mathbf{W}}^{(k)}\left\{\mathbf{\Phi}_{yy}^{(k)} - \mathrm{diag}\left[\mathbf{\Phi}_{yy}^{(k)}\right]\right\}\mathrm{diag}^{-1}\left[\mathbf{\Phi}_{yy}^{(k)}\right], \quad (9)$$

where the superscript $(k)$ is the index of the frequency bin. The demixing matrix should be transferred back to time domain and all $w_{qq,l}$ with $l > L−1$ are set to zero at each iteration to avoid circular convolutive effects.

### III. THE PROPOSED METHOD

When the sources are not continuously mixed, the PSD matrix $\mathbf{\Phi}$ estimation is biased during the time interval when only one source is active or both sources are silent. This will mislead the convergence behavior, resulting in deteriorated separation performance especially when the sources are considerably sparsely mixed.

Our solution to this problem is to execute the offline TRINICON method twice and regularize the update process of the second execution by a straightforward multi-source activity detection based on the preliminary results of separation. Obviously the detection of speech mixtures is a very difficult task. Unlike voice activity detection (VAD) algorithms based on statistics model or spectral characteristic [14-15], the proposed source activity detection is based on the comparison of signal power before and after the preliminary separation process.

The power of the input signal and the output signals of block $m$ are computed as

$$E_x(m) = \frac{1}{2(k_u - k_l + 1)} \sum_{k=k_l}^{k_u} \left( \left| \underline{\mathbf{X}}_1^{(k)}(m) \right|^2 + \left| \underline{\mathbf{X}}_2^{(k)}(m) \right|^2 \right) \quad (10)$$

and

$$E_{yp}(m) = \frac{1}{(k_u - k_l + 1)} \sum_{k=k_l}^{k_u} \left| \underline{\mathbf{Y}}_p^{(k)}(m) \right|^2, \quad (11)$$

where $k_u$ and $k_l$ are the upper bound and the lower bound of the frequency bins. Usually the separation performance in low frequency range is significantly worse than that in middle and high frequency range, therefore the low frequency range signal is not included in the signal power estimation.

When the power of the input signal is lower than the threshold $E_{min}$, it can be determined that both sources are inactive. The power threshold $E_{min}$ is set as $E_{min} = \alpha E_{noise}$, where the background noise power $E_{noise}$ is estimated exploiting methods such as the minima controlled recursive averaging (MCRA) [16-17] and $\alpha$ is a coefficient greater than 1.

Assuming that the component from the $p$th source is suppressed in the $p$th output signal after the preliminary separation process, the power of the $p$th output is expected to be significantly lower than the other output when only the $p$th source is active. Thus if only one output power $E_{yp}$ is obviously lower than input power, the $p$th source can be determined as active. In other cases, it is determined that both sources are active.

The process of the multi-source activity detection is illustrated in Fig. 2, based on which the gradient in (9) for the second separation process is set as

$$\nabla^{NG}_{\underline{\mathbf{w}}^{(k)}} \mathcal{J}^{(k)} = 2 \sum_{i=0}^{\infty} \underline{\mathbf{W}}^{(k)} \left\{ \mathbf{\Phi}^{(k)}_{yy} - \mathrm{diag}\left[\mathbf{\Phi}^{(k)}_{yy}\right] \right\} \mathrm{diag}^{-1}\left[\mathbf{\Phi}^{(k)}_{yy}\right] \mathbf{B}(m), \quad (12)$$

where

$$\mathbf{B}(m) = \begin{bmatrix} \dfrac{\varepsilon_1(m)}{N_{sig}} & 0 \\ 0 & \dfrac{\varepsilon_2(m)}{N_{sig}} \end{bmatrix}, \quad (13)$$

with $N_{sig}$ the number of the whole signal blocks and $\varepsilon_p(m)$ the modified weight whose value is 1 when the $p$th source is active and 0 otherwise.

Note that the precise determination of the activity state for each time segment is impossible. However, the objective of the proposed method is to find enough active speech segments for the filter updating, so the multi-source activity detection shown in Fig. 2 can be regularized in a way to identify more confident active segments. Our numerous simulations demonstrate that significant separation improvement can be achieved when the two sources have similar signal power. In the worst case where the two sources have a large power disparity, the proposed method leads to the performance similar to that of the normal TRINICON framework.

As discussed in Section II, one noncausal filter $w_{12}$ or $w_{21}$ is required when the two sources are in the same half plane as shown in Fig. 1(b). For the time-domain implementation, the filter $w_{qq,l}$ is recommended to be a shifted unit impulse whose shift is set to $L/2$ while exploiting Sylvester constraint computing the $L$th row ($SC_R$, see [18]). However, our numerous experiments and simulations demonstrate that the $L/2$ shift usually does not result in the optimal performance when two sources are not continuously mixed. The shift of the algorithm is tuned to be slightly above the maximum possible delay between two microphones to obtain a better performance.

## IV. SIMULATIONS

The performance of the normal TRINICON and the proposed TRINICON with the multi-source activity detection are compared in simulations. In all the simulations, the sampling frequency is set as $f_s$ = 16 kHz. The microphone signals are obtained by convolving the speech data with room impulse responses (RIRs) generated by the image model [19] with a room of size 6 m × 6 m × 4 m and a varying reverberation time from 150 ms to 350 ms. The dual-channel microphone array with an inter-element spacing of 10 cm is located close to the center of the room. The speakers are 1.5 meter away from the center of the array with different incident angles. The signal power of the two sources are set at the same level resulting in an input signal to interference ratio (SIR) of 0 dB. The active period of each source occupies about 60% of the whole signal respectively, while both sources are active in one third of the duration. The white noise with 30 dB lower power than the speech signals is also added. The shift of the unit impulse in initialization is set to 10 taps.

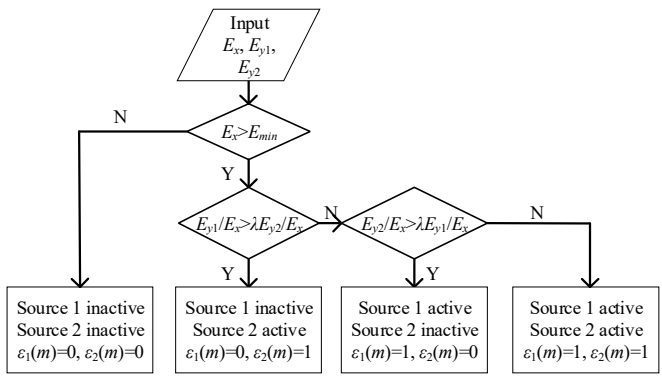

Fig. 2. Procedure of the multi-source activity detection

TABLE I. SIR AND SDR SCORE OF INPUTS AND OUTPUTS UNDER DIFFERENT CONDITIONS

| RT [ms] | DOA of sources [°] | | SIR [dB] | | | SDR [dB] | | |
|---|---|---|---|---|---|---|---|---|
| | source 1 | source 2 | without AD | with AD | Improvement | Input | without AD | with AD |
| 150 | -70 | -15 | 16.4 | 26.1 | 9.7 | 0.2 | 9.4 | 14.5 |
| | | 0 | 24.1 | 26.2 | 2.1 | 0.2 | 16.2 | 15.4 |
| | | 45 | 21.5 | 23.0 | 1.5 | 0.4 | 15.2 | 15.0 |
| | -45 | -15 | 8.7 | 23.8 | 15.1 | 0.2 | 1.3 | 14.4 |
| | | 0 | 16.8 | 25.8 | 9.0 | 0.2 | 10.5 | 15.1 |
| | | 45 | 21.2 | 22.3 | 1.0 | 0.4 | 15.0 | 13.5 |
| 250 | -70 | -15 | 7.8 | 12.8 | 5.0 | -0.6 | 4.6 | 5.8 |
| | | 0 | 11.1 | 16.4 | 5.3 | -0.5 | 5.7 | 7.1 |
| | | 45 | 13.4 | 16.9 | 3.5 | -0.5 | 8.7 | 7.7 |
| | -45 | -15 | 0.5 | 5.0 | 4.5 | -0.5 | -2.4 | 2.4 |
| | | 0 | 5.6 | 14.2 | 8.6 | -0.5 | 3.7 | 6.3 |
| | | 45 | 11.8 | 16.4 | 4.5 | -0.4 | 7.6 | 7.0 |
| 350 | -70 | -15 | 2.5 | 7.1 | 4.7 | -1.6 | -0.9 | 1.9 |
| | | 0 | 2.6 | 12.8 | 10.2 | -1.5 | 0.1 | 4.6 |
| | | 45 | 7.3 | 12.9 | 5.6 | -1.6 | 3.1 | 4.4 |
| | -45 | -15 | 0.2 | 0.6 | 0.4 | -1.5 | -2.8 | -2.1 |
| | | 0 | 1.1 | 5.7 | 4.5 | -1.6 | -1.0 | 1.9 |
| | | 45 | 4.9 | 10.9 | 5.9 | -1.7 | 1.6 | 3.8 |

The BSS_EVAL toolkit [20] is utilized for objective evaluation, and the SIR score and signal to distortion ratio (SDR) score are calculated for each output signal. The higher the SIR value, the more the interfering signal is suppressed. The higher the SDR score is, the less the signals are distorted from the pure speech.

Fig. 3 presents the comparison between the spatial response of the demixing system **W** optimized by the algorithm without and with the multi-source activity detection as the sound sources are located at −70° and −15° with 0.2 s reverberation time. It can be seen that the directivity of the TRINICON without the multi-source activity detection is significantly biased due to the influence of updating in the silent and one-source-active time segments. The improvement of the proposed method with the multi-source activity detection is obvious as the spatial valleys towards the source directions can be clearly seen.

Table I depicts the average SIR and SDR scores of the two output channels over four repeated simulations under different conditions. The improvement of the TRINICON with the multi-source activity detection is validated in most cases, especially when the two sources are in the same half plane with low reverberation time. When the reverberation time increases to 350 ms, the performance of both methods deteriorate, but the advantage of the TRINICON with the multi-source activity detection is still obvious except in the case that both algorithms hardly work with the two sources at −45° and −15° respectively. It should be noted that when the two sources are located in different half planes with large angle disparity, the normal TRINICON behaves well and there is hardly any difference between the two methods.

## V. CONCLUSIONS

A dual-channel BSS system with the frequency domain offline TRINICON framework is investigated and a multi-source activity detection is introduced into the system to improve the separation performance for sparsely mixed signals. The multi-source activity detection is based on the preliminary results of the first execution of the TRINICON, then the source detection results are utilized to regularize the second execution of the TRINICON. Simulations under different acoustics scenarios demonstrate that the proposed system achieves better separation performance for the sparsely mixed signals in most cases.

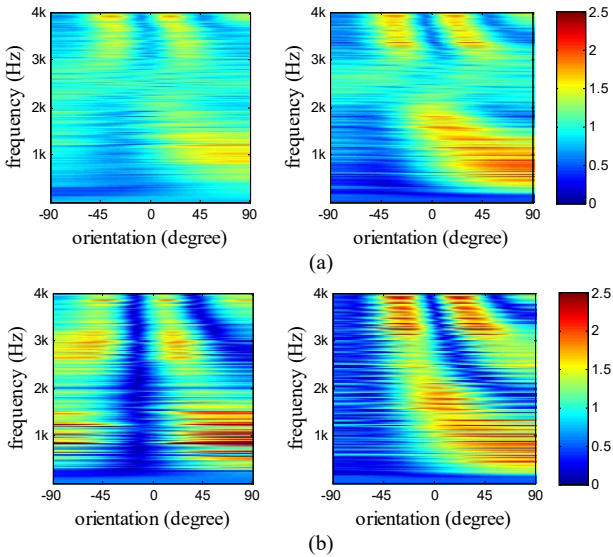

Fig. 3. Spatial response of the optimized demixing system: (a) without and (b) with a multi-source activity detection